\documentclass[a4paper,oneside,final,12pt]{article}
\usepackage{ulem}
\usepackage[english]{babel}
\usepackage{amsmath,amssymb,amsthm}
\usepackage{physics}
\usepackage{setspace}
\usepackage{cite}

\usepackage[warn]{mathtext}
\usepackage{graphicx}
\usepackage{svg}
\usepackage{floatrow}
\usepackage{multirow}
\DeclareGraphicsExtensions{.eps,.png,.jpg}
\usepackage{wasysym}
\usepackage{geometry}
\usepackage{eucal}
\usepackage{indentfirst} 
\usepackage{setspace}
\usepackage{subcaption}
\usepackage{titlesec}

\newcounter{thm}
\newcounter{ex}
\newcounter{re}
\usepackage{amsfonts}

\usepackage{tocloft} 
\usepackage{hyperref}

\begin{document}
\graphicspath{ {picture/} }

\begin{titlepage}
\title{Vacuum Polarization Effects During the Reheating Epoch}
%\title{On the Influence of Vacuum Polarization Effects During the Universe's Reheating Epoch}
\author{A.B. Arbuzov$^{a,b}$, A.A. Nikitenko$^{a}$}
\maketitle
\begin{center}
$^a${Bogoliubov Laboratory of Theoretical Physics, Joint Institute for Nuclear Research, \\
Joliot-Curie st.\,6, Dubna, Moscow region, 141980 Russia} \\
$^b${Dubna State University, Dubna, 141982, Russia}
\end{center}
\begin{abstract}
Quantum effects in the era of reheating of the universe after inflation are considered. A semiclassical approach to gravity theory is applied within Starobinsky's inflationary model with backreaction. Some subtleties associated with accounting for quantum effects in the one-loop approximation are clarified. An estimate of the contribution of vacuum polarization in the stress-energy tensor of matter fields to the cosmological particle creation due to the scalaron decay is presented. 
\end{abstract}
\end{titlepage}

\section{Introduction \label{introdaction}}

In the late 1960s, fairly effective methods for calculating renormalized average values of the stress-energy tensor operator $\bra{\Psi}T_{\mu\nu}\ket{\Psi}$ were developed, allowing these results to be used to study quantum effects in cosmology.
The importance of the stress-energy tensor stems from its position on the right-hand side of Einstein's equations, which allows for a self-consistent description of the interactions of matter, space, and time. 
Note that below we will denote the expectation value of the stress-energy tensor operator $\bra{\Psi}T_{\mu\nu}\ket{\Psi}$ as $\langle T_{\mu\nu}\rangle$, implying unless otherwise specified that the average is taken over the vacuum state~\cite{Grib:1980aih}. 

Quantum effects and especially vacuum polarization effects in cosmology have been actively studied since the pioneering works~\cite{Ginzburg:1971nw,Gurovich:1979xg,Starobinsky:1979ty,Starobinsky:1980te,Vilenkin:1985md}. 
In ref.~\cite{Starobinsky:1980te} a nonsingular solution of so-called semiclassical Einstein equations was obtained, allowing construction of the first inflationary model. Independently, a model of inflation with a scalar (Higgs-like) field was proposed in~\cite{Kazanas:1980tx}. 
Since the metric of homogeneous isotropic cosmology contains only one unknown function, instead of the entire system of gravitational field equations, one equation for its trace can be considered.
When accounting for quantum effects in the case of fields possessing conformally-invariant action at the classical level, the so-called conformal anomaly often arises. 
It manifests as a non-vanishing trace of the stress-energy tensor. Consequently, the anomalous trace often plays an important role when considering cosmological models of the early universe.
One of the first papers in the context of cosmology where the conformal anomaly was discussed was~\cite{Davies:1977ti}. 

In early 1980s, several inflationary models were constructed based on quantum corrections arising from modified Einstein equations with vacuum polarization stress-energy tensor (usually its trace) as source. 
However, it was later understood that inflation can be described using the Starobinsky modified gravity model with the $R^{2}/6M^{2}_{R}$ term, making consideration of vacuum polarization stress-energy tensor redundant.
As became clear later, $f(R)$-type modified gravity models can be reduced to Einstein-Hilbert action plus an additional scalar field called {\em scalaron} with a certain potential. 
Thus, the currently relevant cosmological Starobinsky model can be described either by action
\begin{eqnarray}
    \label{Action_of_Starobinsky}
    S_{tot}= - \frac{M^{2}_{Pl}}{16\pi}\int d^{4}x\sqrt{-g}\left(R - \frac{R^{2}}{6M^{2}_{R}}\right) + S_{\mathrm{matter}},
\end{eqnarray}
or by the action with a scalar field potential, see e.g.~\cite{Nojiri:2004pf,Nojiri:2005sx} and the recent work~\cite{Dorsch:2024nan} devoted to the description of particle production in this model. It should be noted that the consideration of the backreaction associated with the vacuum polarization during the reheating period within the framework of the Starobinsky model was considered in work~\cite{Suen:1987gu}. However, the scalaron decay was not treated there. The magnitude of the correction induced by the vacuum polarization into the value of the scalaron decay width and the corresponding backreaction were not estimated. In this work, we address these issues in more detail.

At the present time, there are many different models of inflation. Very accurate astrophysical data on CMB and matter density distributions allow discrimination of the models, see~\cite{ACT:2025tim} for a recent review. In particular, data on the scalar and tensor primordial perturbation power spectra disfavor the Starobinsky model at the two-sigma level. One of the modern and quite flexible approaches to describing inflation is the alpha-attractor model~\cite{Kallosh:2013tua,Kallosh:2013yoa}, which includes the Starobinsky model as a special limiting case. Nevertheless in this paper, we limit our consideration to the Starobinsky model, as it most naturally accounts for vacuum polarization effects. We expect that our results will be qualitatively applicable to a broader class of inflationary models.

In works~\cite{Arbuzova:2011fu,Arbuzova:2021etq,Arbuzova:2021oqa} the scalaron decay and matter production during the reheating epoch were considered within the Starobinsky cosmological model based on action~\eqref{Action_of_Starobinsky}. 
However, the authors did not account for quantum effects and particularly for vacuum polarization effects on scalaron decay width and related processes. 
Moreover, they mentioned that it would be interesting for future works to add to Starobinsky action a term proportional to the Ricci tensor $R_{\mu\nu}R^{\mu\nu}$ and study how their results would change. 
These questions and some other subtleties, superficially discussed in~\cite{Arbuzova:2011fu,Arbuzova:2021etq,Arbuzova:2021oqa}, are addressed in this work.

The paper is organized as follows. First, in Section~\ref{Discussion_of_gravitational_action_in_relation_to_the_cosmological_model} we briefly review semiclassical gravity and its main properties, focusing on aspects important for cosmological models. 
We clarify the role and magnitude of various effects during inflation and reheating. 
Next, in Section~\ref{Scalar_decay_width_without_taking_into_account_vacuum_polarization_effects} we briefly recall results by Arbuzova and Dolgov~\cite{Arbuzova:2011fu,Arbuzova:2021etq,Arbuzova:2021oqa} on the derivation of the scalaron decay width.
Finally, in Section~\ref{Scalarnon_decay_width_taking__account_vacuum_polarization effects} we study the contribution of the renormalized stress-energy tensor, describing vacuum polarization effects, to scalaron decay width at the leading order. 
We also introduce a procedure allowing conditional separation of oscillation damping during the universe's reheating epoch related to scalaron decay from damping caused by other processes. 
This procedure was implicitly used already in~\cite{Arbuzova:2011fu,Arbuzova:2021etq,Arbuzova:2021oqa} without proper justification. 
We believe that this procedure plays a crucial role in their method and requires a more thorough analysis, which we present at the beginning of this Section.

Within the article, we use the metric signature $(+ - - -)$. Four-dimensional coordinate indices are denoted by Greek letters and run as $0,\ldots,3$. 
We follow the sign conventions for the curvature tensor, the scalar curvature, and the Ricci tensor adopted in~\cite{Landau:1975pou} and in the original work~\cite{Starobinsky:1980te}, which is convenient since authors~\cite{Arbuzova:2011fu,Arbuzova:2021etq,Arbuzova:2021oqa} whose research we develop follow just these conventions. We note that throughout the paper we will work in the Jordan frame, since it is convenient in semiclassical gravity and will allow us to directly compare our results with those obtained previously without taking into account vacuum polarization.

\section{Discussion of Gravitational Action in Relation to the Cosmological Model 
\label{Discussion_of_gravitational_action_in_relation_to_the_cosmological_model}}

Since attempts to quantize gravity from scratch led to complex fundamental problems, physicists decided first to develop quantum field theory in curved spacetime. Specifically, the same fields that had been successfully quantized on the Minkowski background were considered. As a result, it was found that divergent parts, after extraction, turn out to be proportional to the square of scalar curvature and to the full contraction of the Ricci tensor with itself. Thus, the bare Lagrangian before renormalization takes the form~\cite{Grib:1980aih}
\begin{equation}
\label{Action_for_gravity_with_countermembers}   
S = -\frac{M^{2}_{Pl}}{16 \pi}\int d^{4}x\sqrt{-g}\left[R - 2\Lambda + \alpha R^{2} + \beta R_{\mu\nu}R^{\mu\nu}\right].
\end{equation}

It should be noted that the above action does not include all additional terms that must be present in the bare Lagrangian. Namely, terms proportional to $R_{\alpha\beta\gamma\delta}R^{\alpha\beta\gamma\delta}$ and $\Box R$ should be added, but in a four-dimensional spacetime they can be transformed into surface terms using the Gauss–Bonnet formula and thus their variation gives no contribution to equations of motion. For this reason, we omitted them in action~\eqref{Action_for_gravity_with_countermembers}. More details can be found in the well-known monographs~\cite{Grib:1980aih,Birrell:1982ix,Buchbinder:1992rb}, as well as in recent articles, e.g., in ref.~\cite{Matsui:2019tlf}.

Variations of $R^2$ and $R_{\mu\nu}R^{\mu\nu}$ give two new second-rank tensors denoted in the literature~\cite{Grib:1980aih,Birrell:1982ix,Buchbinder:1992rb} as ${}^{(1)}\! H_{\mu\nu}$ and ${}^{(2)}\! H_{\mu\nu}$. 
Their explicit expressions are~\cite{Ford:1997hb,Kiefer:2004xyv}
\begin{equation}
\label{Tensor_H_1_mu_nu}   
{}^{(1)}\! H_{\mu\nu} = \frac{1}{\sqrt{-g}}\frac{\delta \left[\sqrt{-g}R^{2}\right]}{\delta g^{\mu\nu}} = 2\nabla_\nu \nabla_\mu (-R) - 2g_{\mu\nu}\nabla_\rho \nabla^\rho (-R)
 - \frac{1}{2}g_{\mu\nu} R^2 + 2R R_{\mu\nu},
\end{equation}
\begin{eqnarray}
\label{Tensor_H_2_mu_nu}   
{}^{(2)}\! H_{\mu\nu} &\equiv& 
\frac{1}{\sqrt{-g}} \frac{\delta}{\delta g^{\mu\nu}} 
\left[\sqrt{-g} R_{\alpha\beta}R^{\alpha\beta} \right] 
= 2\nabla_\alpha \nabla_\nu (-R_\mu^\alpha) - \nabla_\rho \nabla^\rho (-R_{\mu\nu})
\nonumber \\ &{}& -\frac{1}{2}g_{\mu\nu}\nabla_\rho \nabla^\rho (-R)
-\frac{1}{2}g_{\mu\nu} R_{\alpha\beta}R^{\alpha\beta} + 2R_\mu^\rho R_{\rho\nu},
\end{eqnarray}
where the scalar curvature and the Ricci tensor are taken with opposite signs due to our sign conventions.

Constants $\alpha$ and $\beta$ in counterterms in \eqref{Action_for_gravity_with_countermembers} contain, generally speaking, both infinite parts, used for subtractions and renormalization of the stress-energy tensor of quantum fields, as well as finite parts that can only be determined from experiments or the full quantum gravity (not yet established). Thus, even without quantizing gravity itself, merely considering quantum matter fields on a curved background leads us to a modified gravitational action with curvature-squared terms.

Therefore, in our opinion, the most natural motivation for considering curvature-quadratic modified gravity theories comes from the finite parts of constants $\alpha$ and $\beta$ in counterterms $\alpha R^{2}$ and $\beta R_{\mu\nu}R^{\mu\nu}$, which generally do not vanish. We denote these finite parts as $\alpha_{0}$ and $\beta_{0}$.

Generally, tensor \eqref{Tensor_H_2_mu_nu} is not proportional to \eqref{Tensor_H_1_mu_nu} and must be considered in calculations. However, in cosmology described by the conformally flat metric
\begin{eqnarray}
\label{metrica_Friedmann}   
ds^{2} = dt^{2} - a^{2}(t)\delta_{ij}dx^{i}dx^{j},
\end{eqnarray}
the term $\beta R_{\mu\nu}R^{\mu\nu}$ can be eliminated since its variation satisfies
\begin{eqnarray}
\label{Elimination_square_Ricci_tensor_in_metrica_Friedmann}   
\frac{\delta }{\delta g^{\mu\nu}}\int d^{4}x\sqrt{-g}\left[R_{\alpha\beta}R^{\alpha\beta} - \frac{1}{3}R^{2}\right] = 0.
\end{eqnarray}
Thus, for the conformally flat metric used in refs.~\cite{Arbuzova:2011fu,Arbuzova:2021etq,Arbuzova:2021oqa}, the contribution of the $\beta R_{\mu\nu}R^{\mu\nu}$ term is already captured by an appropriate redefinition of the scalaron mass. Just for this reason, in most works on cosmology, the $\beta R_{\mu\nu}R^{\mu\nu}$ term is not considered, and only the Starobinsky modified action~\eqref{Action_of_Starobinsky} is studied. Note that modern astronomical observations indicate that on cosmological scales the universe is indeed described by the conformally flat Friedmann metric~\eqref{metrica_Friedmann}.

After developing methods for computing and renormalizing the stress-energy tensor of quantum fields, it became possible to use these expressions for studying backreaction effects on the metric by substituting this tensor into the right-hand side of Einstein equations
\begin{equation}
\label{Einstein_equations_with_quantum_effects}   
R_{\mu\nu} - \frac{R}{2}g_{\mu\nu} = \frac{8\pi}{M^{2}_{Pl}}\left[\mathring{T}_{\mu\nu} + \langle T_{\mu\nu}\rangle \right],
\end{equation}
where by $\mathring{T}_{\mu\nu}$ we denote the stress-energy tensor of particles produced from the scalaron decay.
However, the finite parts $\alpha_{0}$ and $\beta_{0}$ were usually set to zero. 
In Sect.~\ref{Scalarnon_decay_width_taking__account_vacuum_polarization effects}, 
we will consider a more general case with
\begin{equation}
\label{Einstein_equations_with_quantum_effects1}   
R_{\mu\nu} - \frac{R}{2}g_{\mu\nu} - \frac{1}{6M_R^{2}}{}^{(1)}\! H_{\mu\nu} = \frac{8\pi}{M^{2}_{Pl}}\left[\mathring{T}_{\mu\nu} + \langle T_{\mu\nu}\rangle \right],
\end{equation}
which appears to be relevant for cosmological models described by the conformally flat metric~\eqref{metrica_Friedmann}.

We will study the influence of vacuum polarization effects on the scalaron decay width in Sect.~\ref{Scalarnon_decay_width_taking__account_vacuum_polarization effects}. The stress-energy tensor describing vacuum polarization in cosmology is given by~\cite{Starobinsky:1980te,Vilenkin:1985md,Grib:1980aih}
\begin{eqnarray}
\label{Stress-Energy_Tensor_mu_nu_vacuum_polarization}   
\langle T_{\mu\nu}\rangle{}^{v.p.} = k_{1}{}^{(1)}\! H_{\mu\nu} + k_{3}{}^{(3)}\! H_{\mu\nu},
\end{eqnarray}
where $k_{1}$ and $k_{3}$ are constants. As we will see below, the exact value of $k_{1}$ is not required. Constant $k_{3}$ depends on the number of scalar $N_{S}$, fermionic $N_{F}$, and gauge $N_{G}$ fields and is of interest to us. It reads
\begin{eqnarray}
\label{k_3_for_H_3_mu_nu}   
k_{3} = \frac{1}{2880\pi^{2}}\left(N_{S} + \frac{11}{2}N_{F} + 62N_{G} \right).
\end{eqnarray}
In the following, we will for brevity omit $v.p.$ in $\langle T_{\mu\nu}\rangle{}^{v.p.}$ and write simply $\langle T_{\mu\nu}\rangle$.

Tensor ${}^{(3)}\! H_{\mu\nu}$ arises in the regularization and renormalization procedure, but it does not correspond to any counterterm in the Lagrangian. Its explicit expression is
\begin{eqnarray}
\label{Tensor_H_3_mu_nu}   
{}^{(3)}\! H_{\mu\nu} = R_{\mu}^{\sigma}R_{\nu\sigma} - \frac{2}{3}RR_{\mu\nu} - \frac{1}{2}g_{\mu\nu}R^{\sigma\tau}R_{\sigma\tau} + \frac{1}{4}g_{\mu\nu}R^{2}.
\end{eqnarray}
Note that tensor ${}^{(3)}\! H_{\mu\nu}$ cannot be obtained by varying a polynomial metric action, since in general it is not divergence-free: $\nabla_{\nu} {}^{(3)}\! H_{\mu}{}^{\nu} \ne 0$. The identity $\nabla_{\nu} {}^{(3)}\! H_{\mu}{}^{\nu} = 0$ holds only for the conformally flat metric considered here. Therefore, the presence of ${}^{(3)}\! H_{\mu\nu}$ in the equations cannot be reduced to any scalar-tensor modified gravity theory.

For our purposes, we also need the expression for the contraction ${}^{(1)}\! H_{\mu}^{\mu}$. It reads
\begin{equation}
\label{Tensor_H_1_mu_nu_with_metrica_Friedmann}   
{}^{(1)}\! H_{\mu}^{\mu} = -2\nabla_{\rho} \nabla^{\rho} R - (-2)\cdot 4\nabla_\rho \nabla^\rho R
 - \frac{1}{2}\cdot 4 R^2 + 2R^{2} = 6\nabla_\rho \nabla^\rho R.
\end{equation}

Since effects related to the backreaction of quantum matter fields on the background spacetime geometry are suppressed by the extremely small gravitational constant (inverse square of the Planck mass), in current studies of early universe models the pure Starobinsky gravitational action~\eqref{Action_of_Starobinsky} is typically used. This approach is taken by authors~\cite{Arbuzova:2011fu,Arbuzova:2021etq,Arbuzova:2021oqa}. However, it is not obvious how vacuum polarization would modify the scalaron decay width into other particles. We will address this question below in Sect.~\ref{Scalarnon_decay_width_taking__account_vacuum_polarization effects} by computing explicitly the vacuum polarization contribution to the scalaron decay width.

\section{Scalaron Decay Width Without Taking Into Account Vacuum Polarization Effects 
\label{Scalar_decay_width_without_taking_into_account_vacuum_polarization_effects}}

In this Section, we briefly describe the method used in works~\cite{Arbuzova:2011fu,Arbuzova:2021etq,Arbuzova:2021oqa} for calculating the scalaron decay width. The authors of the works completely neglected quantum vacuum polarization effects when writing the modified gravitational equations and computing the scalaron decay width. The modified gravitational equations they work with have the form
\begin{equation}
\label{Dolgov_and_Arbuzova_modify_Einstein_equations}   
R_{\mu\nu} - \frac{R}{2}g_{\mu\nu} - \frac{1}{6M^{2}}{}^{(1)}\! H_{\mu\nu} = \frac{8\pi}{M^{2}_{Pl}}\mathring{T}_{\mu\nu},
\end{equation}
where the right-hand side contains the stress-energy tensor derived from quantum matter field expectation values, denoted as $\mathring{T}_{\mu\nu}$. For a detailed discussion and justification, we refer readers to the aforementioned papers~\cite{Arbuzova:2011fu,Arbuzova:2021etq,Arbuzova:2021oqa}, while here we reproduce only the main steps and formulas relevant to our calculations.

In the next Section, we aim to determine how their results change if we include the contribution of the quantum fields' stress-energy tensor expectation value~\eqref{Stress-Energy_Tensor_mu_nu_vacuum_polarization} on the right-hand side of~\eqref{Dolgov_and_Arbuzova_modify_Einstein_equations}. In work~\cite{Arbuzova:2011fu}, an equation describing the cosmological evolution of scalar curvature $R(t)$ including the backreaction is derived. Here, the backreaction refers to the modification of geometry due to the stress-energy tensor of particles produced from scalaron decay. This backreaction is calculated from the solution of equation~\eqref{Dolgov_and_Arbuzova_modify_Einstein_equations}, where the right-hand side contains the stress-energy tensor of decay products.

Since metric~\eqref{metrica_Friedmann} is diagonal and contains only one unknown function, from the entire system~\eqref{Dolgov_and_Arbuzova_modify_Einstein_equations}, it suffices to solve one equation either the $00$-component or equivalently for the trace of equations~\eqref{Dolgov_and_Arbuzova_modify_Einstein_equations}. Authors~\cite{Arbuzova:2011fu,Arbuzova:2021etq,Arbuzova:2021oqa} work with the trace. We will follow their approach and also consider the trace.

The action of a massless scalar field minimally coupled to gravity, into which the scalaron decays, takes the form
\begin{equation}
    \label{Action_of_massless_scalar_field}
    S_{\phi} = \frac{1}{2}\int d^{4}x \sqrt{-g}g^{\mu\nu}\partial_{\mu}\phi\partial_{\nu}\phi.
\end{equation}
For simplicity, we consider the scalar field to be massless, since introducing a mass term for it can only give small corrections proportional to the square of the scalar-to-scalaron mass ratio.

For the Friedmann–Lema\^{\i}tre–Robertson–Walker metric~\eqref{metrica_Friedmann}, the equations of motion for the field satisfying action~\eqref{Action_of_massless_scalar_field} become
\begin{equation}
    \label{eq_of_motion_massless_scalar_field}
    \ddot{\phi} + 3H(t)\dot{\phi} - \frac{1}{a^{2}(t)}\Delta\phi = 0.
\end{equation}

The trace of the stress-energy tensor of a massless scalar field described by action~\eqref{Action_of_massless_scalar_field} reads
\begin{eqnarray}
\label{trace_tensor_T_of_Dolgov}
    \mathring{T}_{\mu}^{\mu} = -g^{\mu\nu}\partial_{\mu}\phi\partial_{\nu}\phi.
\end{eqnarray}

We introduce conformal time $\eta$ related to the cosmological time as
\begin{eqnarray}
    \label{conformal_time_and_scale_factor}
    d\eta = \frac{dt}{a(t)}.
\end{eqnarray}
It is also convenient to make the substitution $\chi = a(t)\phi$.

Then computing the trace on both sides of~\eqref{Dolgov_and_Arbuzova_modify_Einstein_equations} with account of~\eqref{trace_tensor_T_of_Dolgov} and expressing $\phi(t)$ through $\chi(\eta)$, we obtain the following equation
\begin{eqnarray}
\label{integro_diff_eq_of_Dolgov_in_conformal_time}
R^{\prime\prime} + 2\frac{a^{\prime}}{a}R + M^{2}_{R}a^{2}R
= \frac{8\pi M^{2}_{R}}{M^{2}_{Pl}}\frac{1}{a^{2}}\left[\chi^{\prime}{}^{2}-(\vec{\nabla}\chi)^2
+\cfrac{a^{\prime}{}^{2}}{a^2}\chi^{2}-\cfrac{a^{\prime}}{a}(\chi\chi^{\prime}+\chi^{\prime}\chi)\right],
\label{R-diprime}     
\end{eqnarray}
where $\chi(\eta)$ denotes the quantum scalar field expectation value. The quantity $M_{R}$ is often called the scalaron mass, since it characterizes the additional scalar degree of freedom that arises in Staroibnsky's modified gravity. Based on observations of the amplitude of scalar perturbations of the cosmic microwave background radiation, its value is approximately $M_{R}=3 \cdot 10^{13}$ GeV.
The scalar curvature is related to the scale factor via the formula
\begin{eqnarray}
    R = -\frac{6a^{\prime\prime}}{a^{3}}.
    \label{R}
\end{eqnarray}

Estimation of terms on the right-hand side of~\eqref{R-diprime} leads to the following integrals~\cite{Arbuzova:2011fu,Arbuzova:2021etq}:
\begin{eqnarray}
    \label{chi_square_I_1}
    \langle \chi^{2} \rangle \sim -\frac{1}{48\pi^{2}}\int\limits_{\eta_{0}}^{\eta} d\eta_{1} \frac{a^{2}(\eta_{1})R(\eta_{1})}{\eta - \eta_{1}},
\end{eqnarray}
\begin{eqnarray}
    \label{deff_chi_square_I_2}
    \left\langle \langle\partial_{\eta}\chi \rangle^{2} - \langle \vec{\nabla}\chi \rangle^2  \right\rangle \sim -\frac{1}{96\pi^{2}}\int\limits_{\eta_{0}}^{\eta} d\eta_{1} \frac{\partial^{2}_{\eta_{1}}\left( a^{2}(\eta_{1})R(\eta_{1}) \right)}{\eta - \eta_{1}},
\end{eqnarray}
\begin{eqnarray}
    \label{chi_deff_chi_square_I_3}
    \langle \chi\partial_{\eta_{1}}\chi + (\partial_{\eta_{1}}\chi)\chi \rangle \sim -\frac{1}{48\pi^{2}}\int\limits_{\eta_{0}}^{\eta} d\eta_{1} \frac{\partial_{\eta_{1}}\left( a^{2}(\eta_{1})R(\eta_{1}) \right)}{\eta - \eta_{1}}.
\end{eqnarray}

In~\cite{Arbuzova:2011fu,Arbuzova:2021etq}, it is shown that the dominant contribution to the stress-energy tensor from scalaron decay products comes from expression~\eqref{deff_chi_square_I_2}. Then, considering this and coming back from the conformal time to the cosmological one, we finally obtain the following integro-differential equation:
\begin{eqnarray}
\label{integro_differential_equation_of_Dolgov}
\ddot{R} + 3H\dot{R} + M^{2}_{R}R= -\frac{M^{2}_{R}}{12\pi M^{2}_{Pl}}\int\limits_{t_{0}}^{t}dt_{1}\frac{\ddot{R}(t_{1})}{t-t_{1}}.
\end{eqnarray}
The above equation has a singularity at $t_1 = t$, creating additional difficulties for numerical solution. This singularity relates to zero-mode oscillations requiring a separate careful consideration. In~\cite{Arbuzova:2011fu}, it is shown that this divergence can be removed by the scalaron mass renormalization. This is not our research topic, and therefore below we only briefly describe how this occurs.

In the case when the scalar curvature $R(t)$ approximately describes an oscillating function of time at sufficiently large times, equation~\eqref{integro_differential_equation_of_Dolgov} can be solved approximately by introducing a constant $\Gamma$ characterizing the scalaron decay width~\cite{Arbuzova:2011fu,Arbuzova:2021etq,Arbuzova:2021oqa}. In these works, the scalaron decay width is determined using the following approximation for $R(t)$:
\begin{eqnarray}
\label{R_asympt_of_Dolgov_with_R_amp}
R(t) = R_{amp}\cos(\omega t + \theta)e^{-\Gamma t/2},
\end{eqnarray}
where $R_{amp}$ is a slowly varying amplitude of oscillations, approximated by a constant.

In~\cite{Arbuzova:2011fu,Arbuzova:2021etq,Arbuzova:2021oqa}, the term $3H\dot{R}$ in equation~\eqref{integro_differential_equation_of_Dolgov} is either omitted without explanation or briefly noted as negligible. In the next Section, we will explain the reason for this and also provide heuristic arguments justifying the choice of $R(t)$ in the form~\eqref{R_asympt_of_Dolgov_with_R_amp}.

Now, we describe the method used in~\cite{Arbuzova:2021etq} for finding the scalaron decay width. Substituting~\eqref{R_asympt_of_Dolgov_with_R_amp} into the left and right sides of~\eqref{integro_differential_equation_of_Dolgov}, computing all derivatives, and assuming that $\Gamma \ll M_{R}$, we obtain
\begin{eqnarray}
\label{integro_differential_equation_of_Dolgov_2}
\left[\left(\omega^{2} - M^{2}_{R}\right) \cos(\omega t + \theta) + \Gamma\omega \sin(\omega t + \theta)\right]e^{-\Gamma t/2} 
\nonumber \\
= \frac{\omega^{2} M^{2}_{R}}{12\pi M^{2}_{Pl}}e^{-\Gamma t/2} \int\limits_{0}^{t-t_{0}}\frac{d\xi}{\xi}\left[\cos(\omega t+\theta)\cos \omega\xi +\sin(\omega t+\theta)\sin \omega\xi \right].
\end{eqnarray}
Here, the substitution of variables $\xi = t - t_{1}$ under the integral sign was performed.

The integral containing sine can be reduced to the Dirichlet integral using the following transformation:
\begin{eqnarray}
\label{integral_Dirichlet}
\int\limits_{\epsilon}^{t-t_{0}}\frac{d\xi}{\xi}\sin\omega \xi = \int\limits_{-\infty}^{\infty}\frac{dt_{1}}{t_{1}}\frac{e^{i t_{1}} - e^{-i t_{1}}}{4i} = \frac{\pi}{2}.
\end{eqnarray}

The integral containing cosine diverges logarithmically. As shown in works~\cite{Arbuzova:2011fu,Arbuzova:2021etq}, this integral contributes to the scalaron mass renormalization and absorbs the divergence inherent in equation~\eqref{integro_differential_equation_of_Dolgov} at $t = t_{1}$:
\begin{eqnarray}
\label{renormalisation_mass_scalaron}
M^{2}_{R} + \frac{M_{R}^{4}}{12\pi M^{2}_{Pl}}\int\limits_{\epsilon}^{t-t_{0}}\frac{d\tau}{\tau}\cos\omega\tau \mapsto M^{2}_{R}.
\end{eqnarray}

Thus, we can consider in the first approximation $\omega^{2} = M^{2}_{R}$, and the scalaron decay width is determined as 
\begin{eqnarray}
\label{scalaron_decay_width_Dolgова}
\Gamma = \frac{M^{3}_{R}}{24\pi M^{2}_{Pl}}.
\end{eqnarray}

In works~\cite{Arbuzova:2011fu,Arbuzova:2021etq,Arbuzova:2021oqa}, it is justified that the behavior of $H(t)$ and $R(t)$ at sufficiently large $t$ is described by the following asymptotics:
\begin{eqnarray}
\label{Hubble_asympt_of_Dolgov}
H(t) = \frac{2}{3t}\left[1 + \sin(M_{R}t + \theta)\right],
\end{eqnarray}
\begin{eqnarray}
\label{R_asympt_of_Dolgov}
R(t) = - \frac{4M_{R}}{t}\cos(M_{R} t + \theta).
\end{eqnarray}

\begin{figure}[H]
    \includegraphics[width=1.0\textwidth]{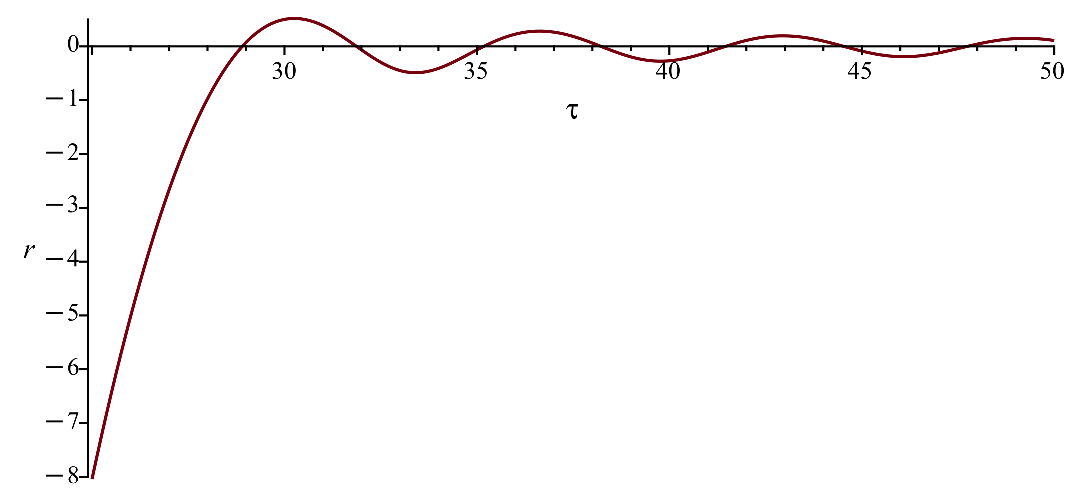}
    \caption{Evolution of the dimensionless scalar curvature $r$ as a function of the dimensionless time $\tau$ for initial values $r(0)=-300$ and $h(0)=5$. }
    \label{f:r_-300_for_scale_8_to_0}   
\end{figure}

\begin{figure}[H]
    \includegraphics[width=1.0\textwidth]{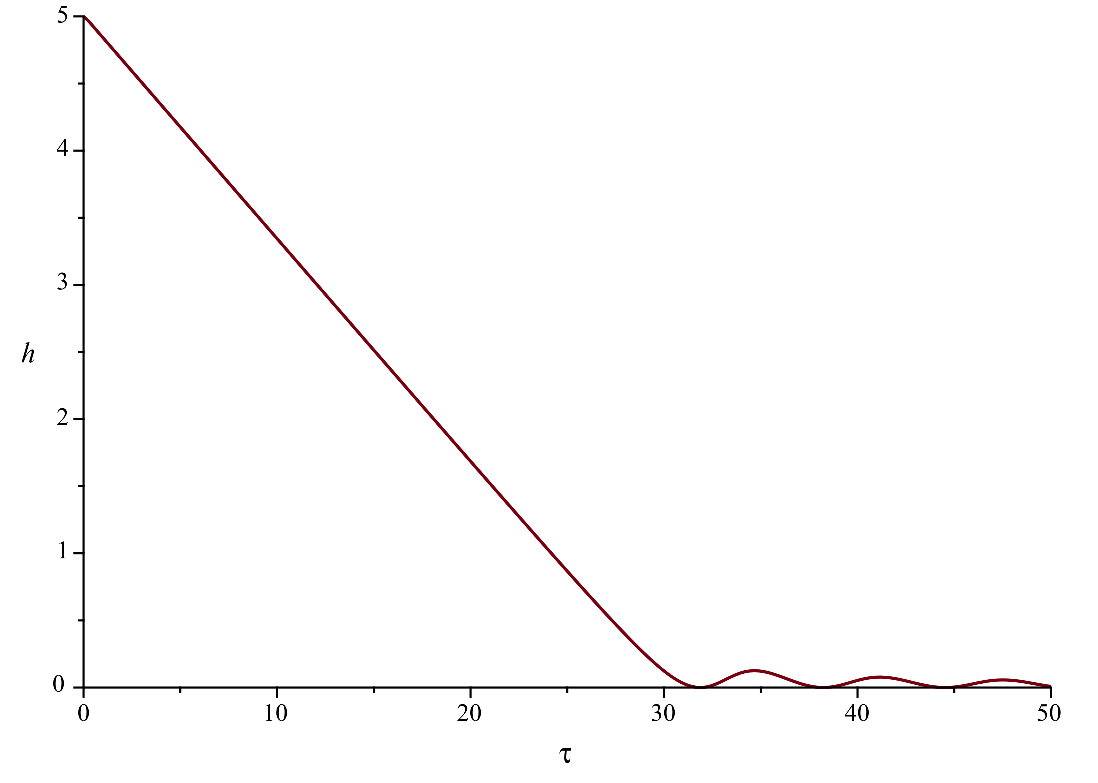}
    \caption{Evolution of the dimensionless Hubble parameter $h$ as a function of the dimensionless time $\tau$ for initial values $r(0)=-300$ and $h(0)=5$.}
    \label{f:h_5}
\end{figure}

Note that in~\cite{Arbuzova:2011fu,Arbuzova:2021etq}, a more general form of these asymptotics is also provided:
\begin{eqnarray}
\label{general_Hubble_asympt_of_Dolgov}
h(\tau) = \frac{h_{0} + h_{1}\sin(\tau + \theta_{h})}{\tau},
\end{eqnarray}
\begin{eqnarray}
\label{general_R_asympt_of_Dolgov}
r(\tau) = \frac{r_{1}\cos(\tau + \theta)}{\tau} + \frac{r_{2}}{\tau^2}.
\end{eqnarray}
It is convenient to introduce the following dimensionless quantities:
\begin{eqnarray}
\label{rel_between_dim and_dimless_quantities}
t=\tau/M_{R}, \ R=M^{2}_{R}r, \ H=M_{R}h, \ \rho=M^{4}_{R}y,
\end{eqnarray}
where $\tau$, $r$, $h$, and $y$ are, respectively, the dimensionless time, the scalar curvature, the Hubble parameter, and the energy density. Note also that in formulas~\eqref{general_Hubble_asympt_of_Dolgov} and \eqref{general_R_asympt_of_Dolgov}, $r(\tau)$ and $h(\tau)$ are dimensionless constants obtained by substituting~\eqref{general_Hubble_asympt_of_Dolgov} and \eqref{general_R_asympt_of_Dolgov} into the dimensionless equations for the trace, the scalar curvature, and the energy density, respectively~\cite{Arbuzova:2021etq}, according to 
\begin{equation}
    \label{equations_Arbuzova_Dolgov}
    \begin{split}
     &r^{\prime\prime}+3hr^{\prime}+r= -8\pi\mu^{2}(1-3w)y \\ 
     &r^{\prime}= -r/6 - 2h^{2} \\ 
     &y^{\prime}+3(1+w)hy=0,
    \end{split}
\end{equation}
and equating terms at the same powers of $1/\tau$. In~\eqref{equations_Arbuzova_Dolgov}, the quantity $w$ denotes the constant in the equation of state $P=w\rho$. Typically, $w$ is chosen as $1/3$ for radiation, $0$ for non-relativistic matter, and $-1$ for vacuum energy (cosmological constant). The numerical solution of equations~\eqref{equations_Arbuzova_Dolgov} with initial values $r(0)=-300$ and $h(0)=5$ is shown in Figs.~\ref{f:r_-300_for_scale_8_to_0} and \ref{f:h_5}. The plots show damped oscillations of the dimensionless scalar curvature  $r(\tau)$ and the dimensionless Hubble parameter $h(\tau)$ for $\tau > 30$. And for $\tau < 30$ one can see the inflation stage.

%\begin{figure}[H]
%    \includegraphics[width=1.0\textwidth]{r_tau_full_new_1}
%    \caption{Evolution of the dimensionless scalar curvature $r$ as a function of dimensionless time $\tau$ for initial values $r(0)=-300$ and $h(0)=5$.}
%    \label{f:r_-300}
%\end{figure} 

\section{Scalaron Decay Width Considering Vacuum Polarization Effects \label{Scalarnon_decay_width_taking__account_vacuum_polarization effects}}

In this Section, we aim to investigate the role of the term $3H\dot{R}$ and vacuum polarization effects in estimating the scalaron decay width using the method outlined in~\cite{Arbuzova:2011fu,Arbuzova:2021etq,Arbuzova:2021oqa}. We will assume that the vacuum polarization is described by the vacuum expectation value of the stress-energy tensor operator \eqref{Stress-Energy_Tensor_mu_nu_vacuum_polarization} on the right-hand side of the Einstein equations.

In works~\cite{Arbuzova:2011fu,Arbuzova:2021etq,Arbuzova:2021oqa}, the term $3H\dot{R}$ was either omitted without explanation or briefly dismissed as negligible. In this study, we deem it necessary to provide clarifications regarding this procedure, as it is non-trivial. Let us consider the equation
\begin{equation}
    \label{eq_for_3H_Gamma_dot_R_with_null_right_side}
\ddot{R} + (3H+\Gamma)\dot{R} + M^{2}_{R}R=0.
\end{equation}
The term $(3H+\Gamma)\dot{R}$ describes the damping of scalaron field oscillations $R(t)$. From the perspective of a clear physical interpretation and computational convenience, we desire to separate the damping associated with the scalaron decay into other particles from the damping caused by other processes. To achieve this, first note that it is assumed $\Gamma\ll M_{R}/\tau$, where $\tau$ represents the dimensionless time at the onset of reheating. Then, the constant $\Gamma$ can be treated as a perturbation to the damping coefficient $3H(t)$, which describes damping unrelated to scalaron decay. Strictly speaking, separating the damping function into a background part (unrelated to scalaron decay) and a perturbation (describing scalaron decay) is impossible, but this can be attempted approximately based on the following reasoning.

First, note that without the $3H$ term in $(3H+\Gamma)\dot{R}$, Equation \eqref{eq_for_3H_Gamma_dot_R_with_null_right_side} reduces to a linear differential equation with constant coefficients (since the decay width is assumed time-independent):
\begin{equation}
    \label{eq_for_R_with_const_dot_R_null_right_side}
    \ddot{R} + \Gamma\dot{R} + M^{2}_{R}R=0,
\end{equation}
whose solution, up to constants, takes the form:
\begin{equation}
\label{integral_for_eq_for_R_with_const_dot_R_null_right_side}
    R(t) \sim \cos(M_{R}t+\theta)e^{-\Gamma t/2}.
\end{equation}
Returning to Equation \eqref{eq_for_3H_Gamma_dot_R_with_null_right_side} but now without $\Gamma$, we have:
\begin{equation}
    \label{eq_for_3Hdot_R_with_null_right_side}
\ddot{R} + 3H\dot{R} + M^{2}_{R}R=0.
\end{equation}   
Since $H(t)$ is a time-dependent function, the solution to this equation cannot be expressed in a convenient analytical form via elementary functions. However, an approximate solution can be obtained under the following simplification. Assume that during reheating, $H(t)$ is described by the asymptotic form derived from \eqref{Hubble_asympt_of_Dolgov}, where the overall factor $2/3t$ in \eqref{Hubble_asympt_of_Dolgov} (multiplied by $\left[1+\sin(M_{R}t+\theta)\right]$) is replaced by $2M_{R}/3\tau$, with $\tau$ being the characteristic dimensionless time at the onset of reheating (assumed to be $\tau \sim 36$). Eq.~\eqref{eq_for_3Hdot_R_with_null_right_side} then becomes
\begin{equation}
    \label{eq_for_3Hdot_R_with_null_right_side_asimp}
\ddot{R} + 2\frac{R_{amp}}{M_{R}}\left[1+\sin(M_{R}t+\theta)\right]\dot{R} + M^{2}_{R}R=0.
\end{equation}
If we discard the oscillatory term $(2R_{amp}/{M_{R}})\sin(M_{R}t+\theta)$ in Eq.~\eqref{eq_for_3Hdot_R_with_null_right_side_asimp}, we obtain
\begin{equation}
    \label{eq_for_3Hdot_R_with_null_right_side_const_coeff}
\ddot{R} + \frac{2R_{amp}}{M_{R}}\dot{R} + M^{2}_{R}R=0,
\end{equation}   
whose integral has the same form as \eqref{eq_for_R_with_const_dot_R_null_right_side}, expressed by the formula
\begin{equation}
    \label{solve_eq_for_3Hdot_R_with_null_right_side_const_coeff}
R(t) \sim \cos\left(t\frac{\sqrt{4M^{2}_{R}-\Gamma^{2}_{0}}}{2}+\theta\right)e^{-\Gamma_{0}t/2}\sim \cos(M_{R}t+\theta)e^{-\Gamma_{0}t/2},
\end{equation}
since \eqref{eq_for_3Hdot_R_with_null_right_side_const_coeff}, like \eqref{eq_for_R_with_const_dot_R_null_right_side}, is an equation with constant coefficients. Here, $\Gamma_{0}$ is introduced as shorthand for $2R_{amp}/M_{R}$. Note that in \eqref{solve_eq_for_3Hdot_R_with_null_right_side_const_coeff}, the final step assumes $\Gamma_{0} \ll M_{R}$.

For $\Gamma_{0}$, we derive
\begin{equation}
    \label{Gamma_0_for_3H_dump}
    \Gamma_{0}=\frac{2M_{R}/t}{M_{R}}=\frac{2M^{2}_{R}/\tau}{M_{R}}=\frac{2M_{R}}{\tau}=1.67 \cdot 10^{12} \ \text{GeV}.
\end{equation}
This quantifies the portion of the coefficient $(3H(t)+\Gamma)$ in $\dot{R}$ that describes damping unrelated to scalaron decay and must be excluded when estimating the scalaron decay width.

We now proceed to evaluate the correction introduced by vacuum polarization induced by the background gravitational field during reheating. The modified gravitational field equations are
\begin{equation}
\label{Modif_Einstein_equations_with_quantum_effects} 
R_{\mu\nu}-\frac{R}{2}g_{\mu\nu}
-\frac{1}{6M_{R}^{2}}{}^{(1)}\! H_{\mu\nu}
=\frac{8\pi}{M^{2}_{Pl}}\left[\mathring{T}_{\mu\nu}
+k_{1}{}^{(1)}\! H_{\mu\nu}+k_{3}{}^{(3)}\! H_{\mu\nu} \right].
\end{equation}
Following the approach adapted in ref.~\cite{Arbuzova:2021etq,Arbuzova:2021oqa,Arbuzova:2011fu}, instead of analyzing the $00$ component of the Einstein equation, we consider the trace of the equation. Taking the trace of Eq.~\eqref{Modif_Einstein_equations_with_quantum_effects} yields
\begin{equation}
\label{Modif_Einstein_equations_with_quantum_effects_1} R-\frac{R}{2}\cdot 4-\frac{1}{6M_{R}^{2}}{}^{(1)}\! H{}^{\mu}_{\mu}=\frac{8\pi}{M^{2}_{Pl}}\left[\mathring{T}^{\mu}_{\mu}+k_{1}{}^{(1)}\! H_{\mu}^{\mu}+k_{3}{}^{(3)}\! H^{\mu}_{\mu} \right],
\end{equation}
\begin{equation}
\label{Modif_Einstein_equations_with_quantum_effects_2} R + \frac{1}{6M^{2}_{R}}{}^{(1)}\! H{}^{\mu}_{\mu}= -\frac{8\pi }{M^{2}_{Pl}}\left[\mathring{T}^{\mu}_{\mu}+k_{1}{}^{(1)}\! H_{\mu}^{\mu}+k_{3}{}^{(3)}\! H^{\mu}_{\mu} \right],
\end{equation}
\begin{equation}
\label{Modif_Einstein_equations_with_quantum_effects_3} R + \left[\frac{1}{6M^{2}_{R}} + \frac{8\pi}{M^{2}_{Pl}}k_{1}\right]{}^{(1)}\! H_{\mu}^{\mu}  = -\frac{8\pi}{M^{2}_{Pl}}\mathring{T}{}^{\mu}_{\mu} - \frac{8\pi}{M^{2}_{Pl}}k_{3}{}^{(3)}\! H^{\mu}_{\mu}.
\end{equation}
Note that in the second term on the left-hand side of Equation \eqref{Modif_Einstein_equations_with_quantum_effects_3}, the term containing $k_{1}$ can be absorbed into the scalaron mass. Thus, we consider the trace equation
\begin{equation}
\label{Trace_equations_Starobinsky_with_quantum_effects_3} R + \frac{1}{6M^{2}_{R}}{}^{(1)}\! H_{\mu}^{\mu}  = -\frac{8\pi}{M^{2}_{Pl}}\mathring{T}{}^{\mu}_{\mu} - \frac{8\pi}{M^{2}_{Pl}}k_{3}{}^{(3)}\! H^{\mu}_{\mu},
\end{equation}
where
\begin{equation}
\label{trace_ensor_particle_creation_T}
    \frac{8\pi}{M^{2}_{Pl}}\mathring{T}{}^{\mu}_{\mu}=\frac{M^{2}_{R}}{12\pi M^{2}_{Pl}}\int\limits_{t_{0}}^{t}dt_{1}\frac{\ddot{R}(t_{1})}{t-t_{1}},
\end{equation}
as in \eqref{integro_differential_equation_of_Dolgov}.
The trace of the tensor ${}^{(1)}\! H_{\mu
u}$ is
\begin{equation}
\label{Trace_tensor_H_1_Friedmann} R + \frac{1}{6M^{2}_{R}}{}^{(1)}\! H{}^{\mu}_{\mu}= \frac{1}{M^{2}_{R}}\nabla_\rho \nabla^\rho R + R=\frac{1}{M^{2}_{R}}\left(\ddot{R}+3H\dot{R}\right) + R,
\end{equation}
using Eq.~\eqref{Tensor_H_1_mu_nu_with_metrica_Friedmann} and the relation between the scale factor and the Hubble parameter. The trace of the tensor ${}^{(3)}\! H_{\mu\nu}$ can also be expressed in terms of the Hubble parameter, which we evaluate using the asymptotic form \eqref{Hubble_asympt_of_Dolgov} for $H(t)$:
\begin{equation}
  \begin{split} 
    \label{eq_for_tensor_H_3_through_H}
      &{}^{(3)}\! H^{\mu}_{\mu}=12\left(H^{2}\dot{H} + H^{4}\right)=12\left[ \left(\frac{2}{3t}\left[1+\sin(M_{R}t+\theta)\right]\right)^{2}\frac{d}{dt}\left[ \frac{2}{3t}\left[1+\sin(M_{R}t+\theta)\right] \right] \right. \\ 
      &\left. + \left(\frac{2}{3t}\left[1+\sin(M_{R}t+\theta)\right]\right)^{4} \right].
    \end{split}
\end{equation}
Our goal is to obtain the leading-order approximation for the scalaron decay width, so we neglect all powers of $\sin(M_{R}t+\theta)$ beyond the first and all other harmonics. Evaluating the right-hand side of Eq.~\eqref{eq_for_tensor_H_3_through_H}, the required term involving ${}^{(3)}\! H^{\mu}_{\mu}$ linear in $\sin(M_{R}t+\theta)$ becomes:
\begin{equation}
  \label{eq_for_tensor_H_3_through_sin}
   \frac{8\pi M^{2}_{R}}{M^{2}_{Pl}}{}^{(3)}\! H^{\mu}_{\mu}12k_{3}\sim \frac{8\pi M^{2}_{R}}{M^{2}_{Pl}}12 k_{3}\left(-\frac{8}{81t^{4}} \right)\sin(M_{R}t+\theta).
\end{equation} 
Collecting terms proportional to $\sin(M_{R}t+\theta)$, we derive
\begin{eqnarray}
    \label{integro_diff_eq_of_Dolgov_with_vacuum_polarization}
    \begin{split}
&R_{amp}\left[\left(\omega^{2} - M^{2}_{R}\right) \cos(\omega t +\theta) + (\Gamma_{0}\omega+\Gamma\omega - 2R_{amp}) \sin(\omega t +\theta)\right]e^{-(\Gamma_{0}+\Gamma) t/2} \sim \\
&R_{amp}\frac{\omega^{2} M^{2}_{R}}{12\pi M^{2}_{Pl}}e^{-(\Gamma_{0}+\Gamma) t/2} \int\limits_{0}^{t-t_{0}}\frac{d\xi}{\xi}\left[\cos(\omega t+\theta)\cos \omega\xi +\sin(\omega t+\theta)\sin \omega\xi \right] \\
& \qquad - \frac{8\pi M^{2}_{R}}{M^{2}_{Pl}}12 k_{3}\left(-\frac{8}{81t^{4}} \right)\sin(\omega t+\theta),
    \end{split}
\end{eqnarray}
where substitution of Eq.~\eqref{Gamma_0_for_3H_dump} for $\Gamma_{0}$, along with $R_{amp}\sim M^{2}_{R}/\tau$ and $\omega^{2}=M^{2}_{R}$, leads to cancellation of $\Gamma_{0}\omega$ and $2R_{amp}$.

We now estimate the correction due to vacuum polarization in the scalaron decay width during the initial reheating period determined by $\tau=36$. The maximum effect is expected at this stage, as the amplitudes of Hubble parameter and scalar curvature oscillations are largest initially and decay over time. The relations between $R_{amp}$ in Eq.~\eqref{R_asympt_of_Dolgov_with_R_amp}, scalaron mass, and time are given by
\begin{equation}
    \label{Characteristic_time_start_reheating}
    \frac{1}{t}=\frac{M_{R}}{\tau}=\frac{M_{R}^{2}}{\tau}\frac{1}{M_{R}}=\frac{R_{amp}}{M_{R}},
\end{equation}
using $\tau=M_{R}t$. To estimate the leading correction from the stress-energy tensor, we express $8/(81t^{4})$ in terms of the scalaron mass and the dimensionless time at the reheating onset:
\begin{equation}
  \label{1_t_pow_4}
    \frac{1}{t^{4}}=\left( \frac{M_{R}}{\tau} \right)^{4}=\frac{M^{2}_{R}R_{amp}}{\tau^{3}}.
\end{equation} 
Since we are interested only in the leading term and neglect higher harmonics, Eq.~\eqref{eq_for_tensor_H_3_through_sin} retains only the relevant terms proportional to $\sin(M_{R}t+\theta)$. The coefficients of $\sin(M_{R}t+\theta)$ in Eq.~\eqref{integro_diff_eq_of_Dolgov_with_vacuum_polarization} satisfy
\begin{equation}
    \label{eq_R_amp_scalar_decay_width_Gamma_0_and_Gamma}
    R_{amp}M_{R}\Gamma e^{-(\Gamma_{0}+\Gamma) t/2}\sim R_{amp}\frac{M^{4}_{R}}{12\pi M_{Pl}^{2}}\frac{\pi}{2}e^{-(\Gamma_{0}+\Gamma) t/2} + \frac{8\pi M^{2}_{R}}{M^{2}_{Pl}}12k_{3}\left(\frac{8}{81t^{4}}\right).
\end{equation}
Note that computing Eq.~\eqref{eq_for_tensor_H_3_through_H} using asymptotics \eqref{R_asympt_of_Dolgov_with_R_amp} and \eqref{Hubble_asympt_of_Dolgov} introduces constants besides $\sin(M_{R}t+\theta)$. However, these constants (e.g., $const/\tau^{2}$ from \eqref{general_R_asympt_of_Dolgov}) do not contribute to our calculations and are omitted.

Multiplying both sides by $e^{-\Gamma_{0} t/2}$ and using $\Gamma_{0}=2M_{R}/\tau$, $t=M_{R}/\tau$, and $e^{\Gamma_{0}t/2}=e$, we derive
\begin{equation}
   \begin{split} 
  \label{eq_with_R_amp_scalar_decay_width_with_vacuum_polarization}
    &R_{amp}\Gamma M_{R}=R_{amp}\frac{M^{4}_{R}}{12\pi M_{Pl}^{2}}\frac{\pi}{2}+\frac{8\pi M^{2}_{R}}{M^{2}_{Pl}}12k_{3}\left(\frac{8}{81t^{4}}\right)e \\
    & = R_{amp}\frac{M^{4}_{R}}{24 M_{Pl}^{2}}+\frac{256\pi M^{2}_{R}}{27M^{2}_{Pl}}k_{3}\left(\frac{M^{2}_{R}R_{amp}}{\tau^{3}}\right)e,
    \end{split} 
\end{equation}
where the Taylor expansion of $e^{-\Gamma t/2}$ retains its leading term.

Finally, the scalaron decay width including vacuum polarization corrections becomes
\begin{equation}
  \label{scalar_decay_width_with_vacuum_polarization}
    \Gamma =\frac{M^{3}_{R}}{24 M_{Pl}^{2}}+\frac{256\pi M^{3}_{R}}{27M^{2}_{Pl}\tau^{3}}k_{3}e,
\end{equation}
with $k_{3} \sim 0.036$ from \eqref{k_3_for_H_3_mu_nu}. The number of scalar, fermionic, and gauge fields included in the stress-energy tensor were taken as $N_{S}=104$, $N_{F}=32$, $N_{G}=12$ \cite{Matsui:2019tlf}.

The first term in Eq.~\eqref{scalar_decay_width_with_vacuum_polarization} represents the scalaron decay width calculated in ~\cite{Arbuzova:2011fu,Arbuzova:2021etq,Arbuzova:2021oqa}, while the second term accounts for vacuum polarization effects at the beginning of reheating. Quantitatively, the vacuum polarization correction evaluates to
\begin{equation}
  \label{correction_vacuum_polarization_for_scalaron_decay_width}
    \Delta\Gamma =\frac{256\pi M^{3}_{R}}{27M^{2}_{Pl}\tau^{3}}k_{3} e \sim \frac{256\pi (3 \cdot 10^{13})^{3} \cdot 0.036}{27 \cdot (1.22 \cdot 10^{19})^{2} \cdot 40^{3}}2.72 \sim 0.01 \ \text{GeV}.
\end{equation}
Recall that the scalaron decay width without vacuum polarization contribution is $\sim 7.56$ GeV. Thus, the vacuum polarization correction causes only a small increase in the decay width. However, its contribution is suppressed not by the scalaron-to-Planck mass ratio but by the cube of the characteristic dimensionless time during reheating. Therefore, the approximation used in ref.~\cite{Arbuzova:2011fu,Arbuzova:2021etq,Arbuzova:2021oqa}, which neglects the vacuum polarization stress-energy tensor, remains valid for describing the scalaron decay width during reheating.
Nevertheless, the considered effect can become relevant in the future with increasing of the accuracy of cosmological observational data. In particular, the contributions of various components of the the universe energy density directly depend on the value of the scalaron decay width.

\section{Conclusion}

The role of various contributions of terms in the quadratic gravity model has been analyzed within the context of homogeneous isotropic cosmology. It has been demonstrated that, in this case, the Starobinsky modified gravity model (or its equivalent formulation as a scalar field theory with a specific potential) represents the simplest non-trivial modification of General Relativity (GR) that is quadratic in curvature for cosmological applications. Practically all modifications of GR motivated by considerations of quantum fields in curved spacetime are reduced to the Starobinsky Lagrangian in the framework of homogeneous and isotropic cosmology.

The role of the term $3H\dot{R}$ in the equation governing the evolution of scalar curvature during the determination of the scalaron decay width has been analyzed. The dominant contribution describing the damping of curvature oscillations unrelated to the scalaron decay has been isolated. It has been shown that this contribution can be approximately characterized by selecting a characteristic time corresponding to the onset of the universe's reheating phase, thereby separating it from the damping effects caused by the scalaron's decay.

The scalaron decay width has been calculated taking into account corrections due to vacuum polarization, 
described by the stress-energy tensor of quantum matter fields. It has been demonstrated that the relative 
magnitude of the correction introduced by vacuum polarization into the scalaron decay width is inversely
proportional to the third power of the dimensionless time parameter at the reheating epoch. Thus, such corrections may become relevant in the future to test various modified gravity models using increasingly accurate observational data. It would be valuable to verify our results by numerically solving the integro-differential equation~\eqref{Trace_equations_Starobinsky_with_quantum_effects_3}. However, this task constitutes a topic for separate future research.

\bibliographystyle{elsarticle-num}
\bibliography{vac_pol_scalaron_width_v2}

\end{document}